\documentclass{isprs} 

\usepackage{subfigure}

\usepackage{setspace}
\usepackage{geometry} 
\usepackage{epstopdf}
\usepackage[labelsep=period]{caption}  
\usepackage[british]{babel} 
\usepackage[hang]{footmisc}
\usepackage{natbib}
\usepackage[T1]{fontenc}
\usepackage{lmodern}
\usepackage[hidelinks]{hyperref}

\begin{document}

\title{Checkerboard Target Measurement in Unordered Point Clouds with Coloured ICP}
\date{}


\author{
 June Moh Goo\thanks{Corresponding author}, Jialun Li, Darmawan Wicaksono, Jan Boehm}

\address{
	Department of Civil, Environmental and Geomatic Engineering, University College London, UK\\
 \cr – \{june.goo.21, jialun.li.23, darmawan.wicaksono.23, j.boehm\}@ucl.ac.uk\\
}



\abstract{

In this work we investigate the problem of measuring a checkerboard target’s centre in an 3D point cloud. This is an important problem which has applications in registration, long term monitoring and linking to other sensor systems. We use a 3D template matching approach based on the coloured ICP algorithm to solve the problem. We tackle the problem under the additional constraints that we assume no structure in the 3D data in order to be able to handle unordered point clouds. This gives us the capability to process data from the new generation of low-cost LIDAR sensors. This category of sensors also suffers from increased noise in range and reflectivity measurement. We provide extensive simulation results using synthetic data to capture the potential of the approach. We then give the detailed steps for handling real sensor data. The code and dataset are available at \href{https://github.com/3Dimaging-ucl/3D_ICP}{https://github.com/3Dimaging-ucl/3D\_ICP}.
}


\maketitle

\sloppy


\section{Introduction}\label{INTRODUCTION}

\begin{figure*}[ht]
\minipage{0.32\textwidth}
  \includegraphics[width=\linewidth]{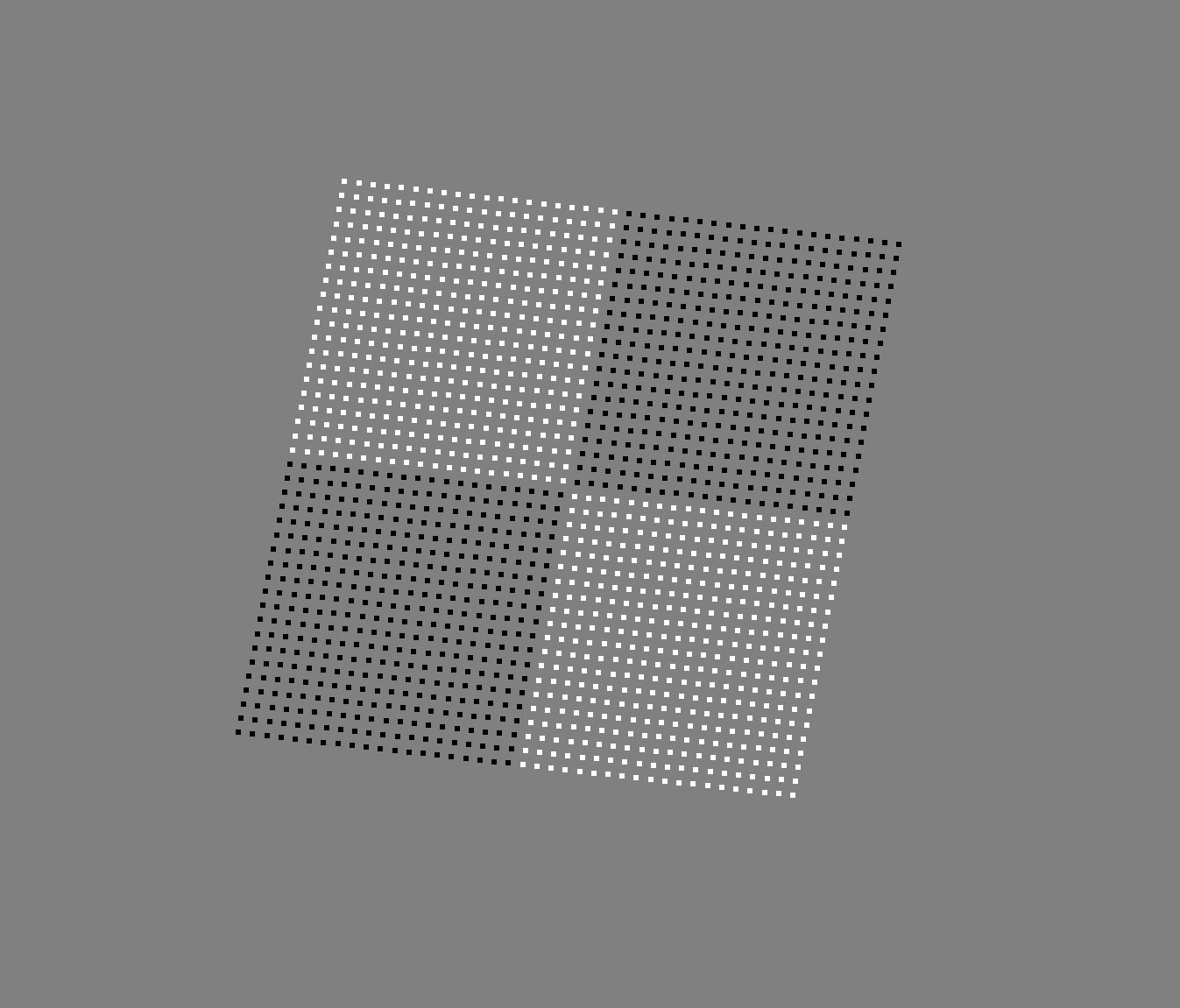}
\endminipage\hfill
\minipage{0.32\textwidth}
  \includegraphics[width=\linewidth]{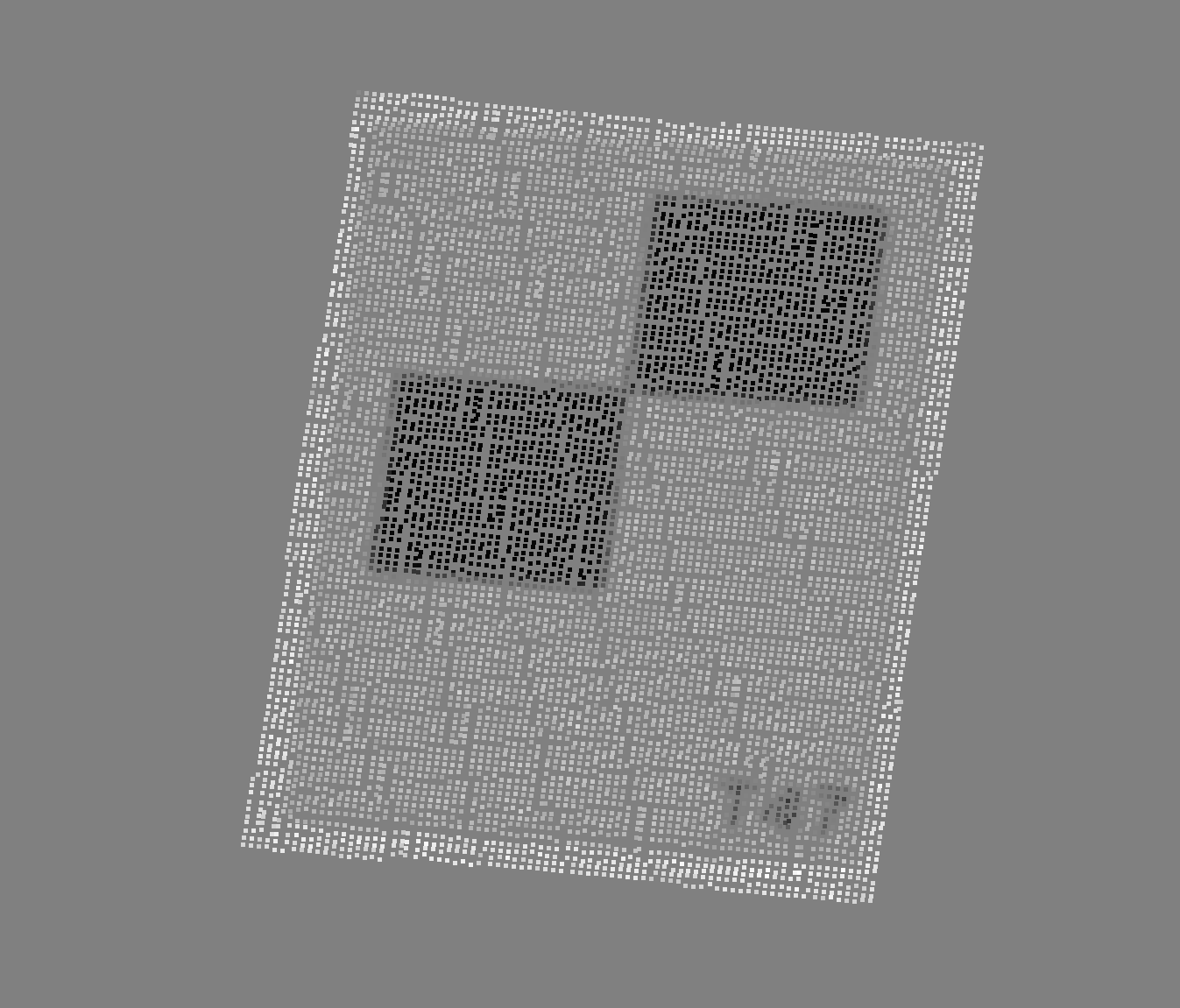}
\endminipage\hfill
\minipage{0.32\textwidth}%
  \includegraphics[width=\linewidth]{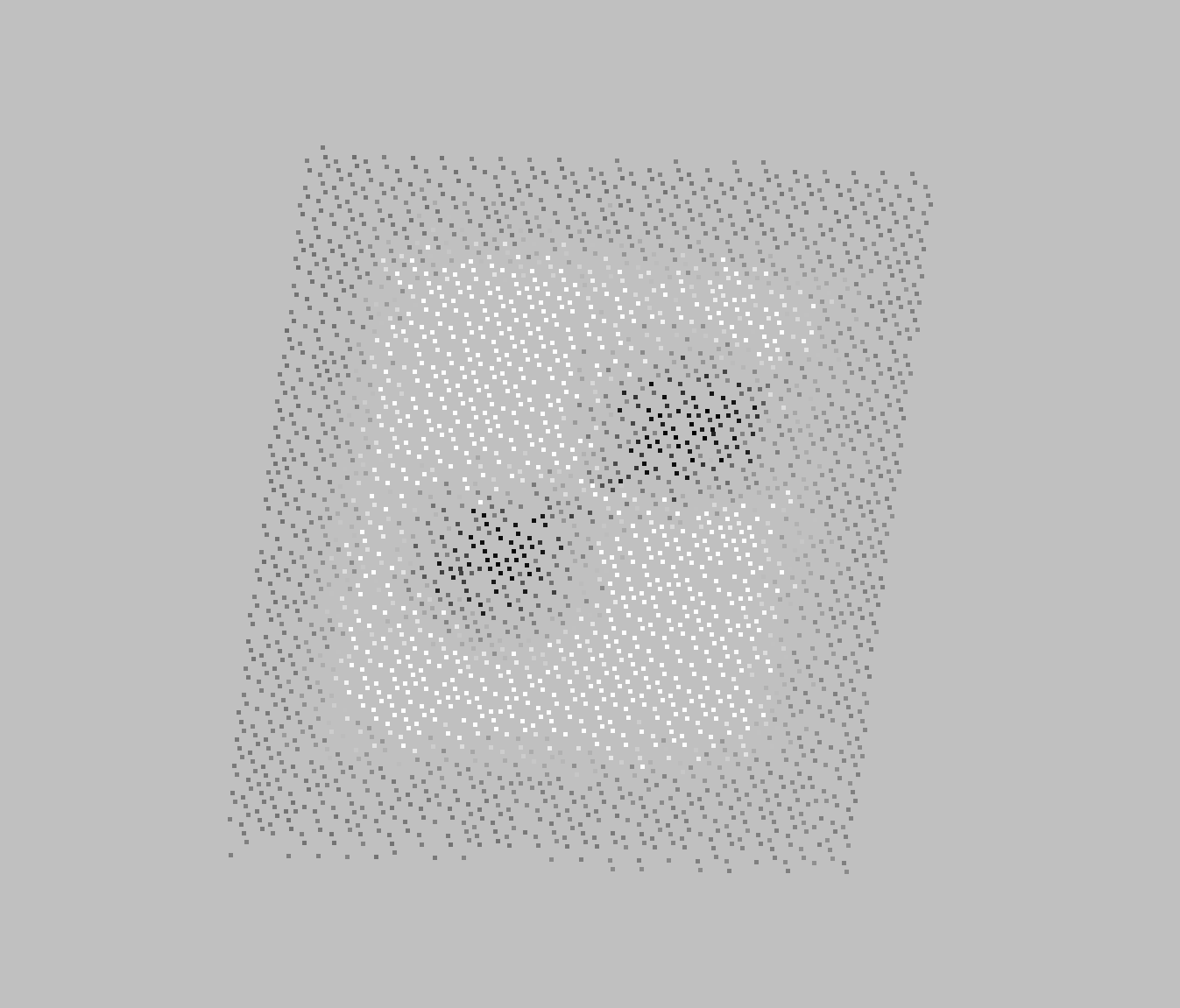}
\endminipage
\caption{Visual comparison of a synthetic target template, survey-grade instrument scan of a target and low-cost scan of a target.}
\label{fig:target_visual}
\end{figure*}

It is a common workflow scenario for terrestrial laser scanning that several separate stations are necessary to capture a scene. The alignment of the stations into a common reference frame is usually referred to as registration. Several alternative approaches exist to solve the alignment problem. Often, approaches for terrestrial laser scanning can be traced back to classical surveying, photogrammetry, or computer vision. We can categorise the approaches into marker based, sensor based or data driven approaches \citep{pfeifer_early_2008}. The marker-based approaches are still expected to deliver the highest accuracy. 

Similar use cases for high accuracy targets arise for easy integration of terrestrial laser scans with other measurement technologies such as total station or laser tracker. Also, the long-term observation of single qualified points for example in monitoring or metrology applications can be directly accomplished with targets. For this reason, commercial laser scanning systems provide vendor specific software solutions to detect markers in laser scans from their systems. Although different target designs exist \citep{jansen_decreasing_2019}, checkerboard style targets have been universally adopted across vendors for reasons of cost and versatility.

Due to the raster-wise sampling of most common terrestrial laser scanning systems the back-scattered intensity can be represented in a matrix structure \citep{bohm_automatic_2007,sanchez_castillo_semantic_2021}. This allows for the adoption of robust and widely available image processing algorithms to perform target detection and measurement in 2D. Unfortunately, for non-standard laser scanners, this is not always possible. Some newer generation scanners deliver unordered point clouds \citep{arteaga2019}. It is not possible to apply image processing algorithms directly to this data type. One possible approach is to project (a portion of) the unordered point cloud into a plane and use a 2D approach in the projection plane \citep{ge_target_2015, goo2024zero}. This would then allow to revert to standard image processing. 

In this work, we want to follow another approach. We formulate the problem of measuring a target's centre as a template matching problem in 3D. We distinguish the measurement problem from the detection problem. Detection tries to find the target in the overall scene. Here we are concerned with finding the exact centre of the target when its approximate location is already known. We solve the template matching directly on the raw 3D point cloud data and thus avoid the projection of the data to 2D described above. 

In unordered 3D data there is no direct equivalent to a cross-correlation. Instead, the Iterative Closest Point (ICP) category of algorithms serves a similar purpose for template matching \citep{besl_method_1992}. It is well-known that ICP does not perform well on planar surfaces. Therefore, additional information needs to be taken into account. Using per-point intensity for ICP is a common approach and has been successfully applied to this problem \citep{liang_fast_2024}. 

Naively, we can extend the 3D geometry information to form a 4D vector by adding intensity \citep{feldmar_extension_1997}. However, the more recent literature of \cite{park2017} suggests a different approach. Two error terms separating geometry and intensity (or colour) are formulated and optimised for. This solution is reported to have superior precision over previous approaches. An open implementation of this algorithm is available in the widely adopted Open3D library \citep{Zhou2018}. Based on this implementation, we perform several experiments both using synthetic and real data to test the precision of the target measurement using a checkerboard target template.

\section{Related Works}\label{RELATED_WORKS}
\begin{figure*}[t]
  \centering
    \includegraphics[width=1\linewidth]{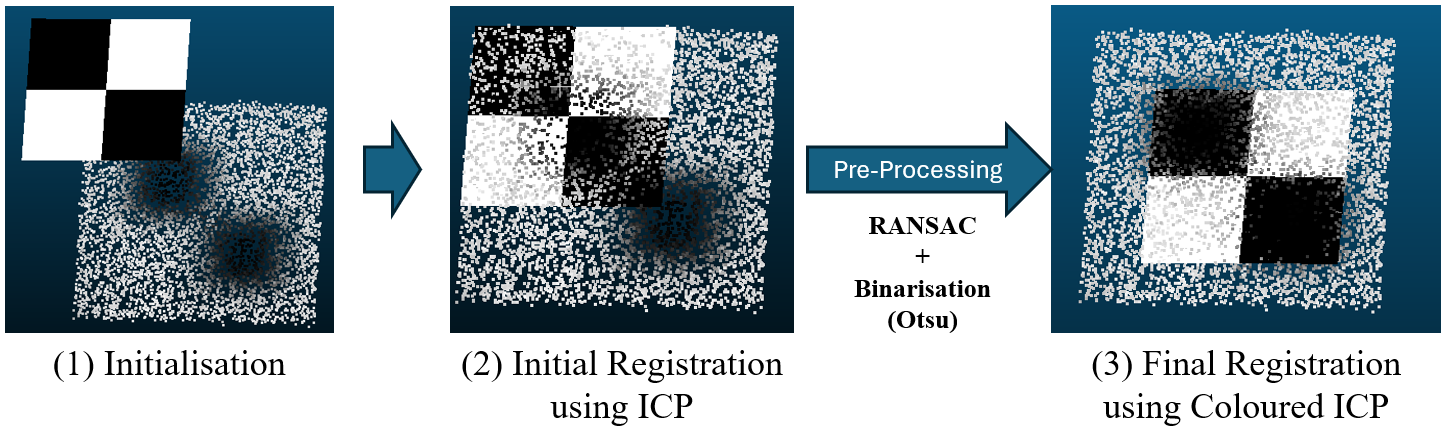}
    \caption{Target measurement process on low-cost scan data using ICP and Coloured ICP. (1) Initialisation: The source point cloud (checkerboard) is misaligned with the target point cloud. (2) Initial Registration using Point-to-Plane ICP: Standard ICP leads to suboptimal registration. (3) Final Registration using Coloured ICP: Colour information is incorporated after pre-processing with RANSAC and Binarisation with Otsu Thresholding for real data, resulting in improved alignment.}
    \label{fig:Registration_visualisation}
\end{figure*}

\subsection{Iterative Closest Point (ICP) Algorithm}
The Iterative Closest Point (ICP) algorithm has been a fundamental technique in 3D computer vision and robotics for point cloud. Originally proposed by \cite{besl_method_1992}, ICP aims to minimise the distance between two datasets, typically referred to as the source and the target. The algorithm operates in an iterative manner, identifying correspondences by matching each source point with its nearest target point \citep{survey_ICP}. It then computes the rigid transformation, usually involving both rotation and translation, to achieve the best alignment of these matched points \citep{survey_ICP}. This process is repeated until convergence, where the change in the alignment parameters or the overall alignment error becomes smaller than a predefined threshold.

One key advantage of the ICP framework lies in its simplicity: the algorithm is conceptually straightforward, and its basic version is relatively easy to implement. However, traditional ICP can be sensitive to local minima, often requiring a good initial alignment \citep{zhang2021fast}. Furthermore, outliers, noise, and partial overlaps between datasets can significantly degrade its performance \citep{zhang2021fast, bouaziz2013sparse}. Over the years, various modifications and improvements \citep{gelfand2005robust, rusu2009fast, aiger20084, gruen2005least, fitzgibbon2003robust} have been proposed to mitigate these issues. Among the most common strategies are robust cost functions \citep{fitzgibbon2003robust}, weighting schemes for correspondences \citep{rusu2009fast}, and more sophisticated techniques \citep{gelfand2005robust, bouaziz2013sparse} to reject outliers. 

In addition, there is significant interest in integrating additional information into the ICP pipeline. Instead of solely relying on geometric cues such as point coordinates or surface normals, recent approaches have proposed incorporating colour (RGB) or intensity data to enhance correspondence accuracy. These methods \citep{park_colored_2017, 5980407}, commonly known as "Colored ICP" employ differences in pixel intensities or colour values as additional constraints. This is particularly beneficial in situations where geometric attributes alone are inadequate for accurate alignment or where surfaces possess complex texture patterns that can assist in the matching process.

\subsection{Applications of Target Measurement}

One approach relies on the use of physical checkerboard targets for registration. \cite{fryskowska2019} analyse checkerboard target identification for terrestrial laser scanning. They propose a geometric method to determine the target centre with higher precision, demonstrating that their approach can reduce errors by up to 6 mm compared to conventional automatic methods.

\cite{becerik2011assessment} examines data acquisition errors in 3D laser scanning for construction by evaluating how different target types (paper, paddle, and sphere) and layouts impact registration accuracy in both indoor and outdoor environments and presents guidelines for optimal target configuration.

\citet{Liang2024} propose to use Coloured ICP to measure target centres for checkerboard targets, similar to our investigation. They use data from a survey-grade terrestrial laser scanner. Their intended application is structural bridge monitoring purposes. They report an average accuracy of the measurement below 1.3 millimetres.

Where targets cannot be placed in the scene, the intensity information form the scanner can still be used to identify distinctive points. For point cloud data that is captured with a regular pattern, standard image processing can be used in a similar way to target detection. For example, \citet{wendt_automation_2004} proposes to use the SUSAN operator on a co-registered image from a camera, \citet{bohm_automatic_2007} proposes to use the SIFT operator on the LIDAR reflectance directly and \citet{theiler_markerless_2013} propose to use a Difference-of-Gaussian approach on the reflectance information.
Most of these methods extract image features to find reliable 3D correspondences for the purpose of registration.

In the following we describe our approach to the measurement of the target centre. In contrast to most proposed methods above we focus on unordered point clouds, where raster-based methods are not available, and low-cost sensors, where increased measurement noise and outliers are expected. As we are not aware of a commercial reference solution to this problem, we start by conducting a series of synthetic experiments to explore the viability and accuracy potential of the approach.


\section{Methodology}\label{METHODOLOGY}


\begin{figure}[h]
    \centering
    \includegraphics[width=1\linewidth]{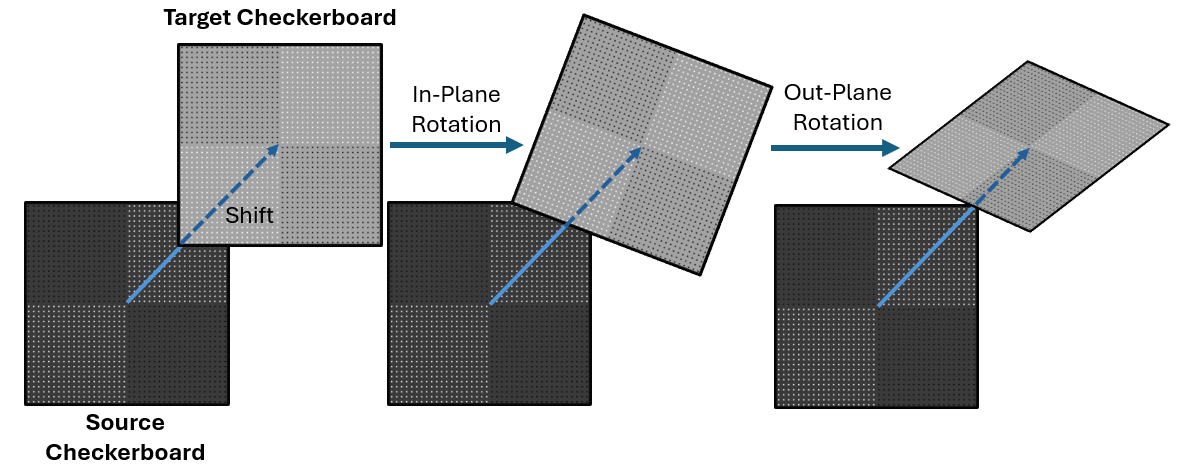}
    \caption{The process of target generation}
    \label{fig:target_generation_process}
\end{figure}
\textbf{Synthetic Data} The initial set of experiments focuses on generating synthetic point cloud data to evaluate the performance of Coloured ICP algorithm. Using synthetic data guarantees precise ground truth values for the target positions and orientations. It also allows us to explore the iterative algorithm under a variety of initial configurations of translation and orientation. We can also control the amount of noise the simulated LIDAR data.
As we vary translation, orientation, and noise our evaluation criteria are: 
\begin{itemize}
    \item \textbf{Root Mean Squared Error (RMSE)}: The error between the source and target checkerboards after alignment. The error is expressed as a fraction of the target size.
    \item \textbf{Success of measurement}: Defined as a case where ICP or Coloured ICP converges with RMSE less than 15\% of one side of the source checkerboard (geometric success) and the colour matching score exceeds 0.5 (colour success), with both conditions satisfied.
\end{itemize}

We vary the initial translation of the template in relative terms of the size of the target from 0\% (full overlap) to 150\% (no overlap) in steps of 10\%, applying both in-plane and out-plane rotations with the same degree of rotation for each (see Figure \ref{fig:target_generation_process}). We conduct experiments by randomly selecting translation directions while increasing the shift and rotation magnitudes. For each configuration, we perform 100 experiments and compute the statistics for the criteria above.

\begin{figure}[h]
    \centering
    \subfigure[Before RANSAC]{
        \includegraphics[width=0.45\linewidth]{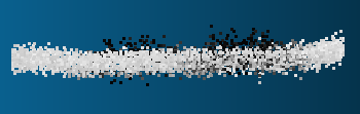}
    }
    \hfill
    \subfigure[After RANSAC]{
        \includegraphics[width=0.45\linewidth]{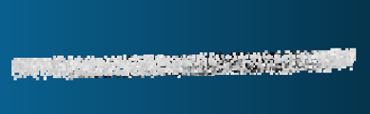}
    }
    \caption{Effect of RANSAC filtering. (a) Before RANSAC: Numerous outliers are present particularly in the black regions, affecting the alignment. (b) After RANSAC: The outliers are removed, leading to a cleaner point cloud and improved registration quality.}
    \label{fig:ransac_before_after}
\end{figure}

\begin{figure}[h]
    \centering
    \subfigure[Before binarisation]{
        \includegraphics[width=0.45\linewidth]{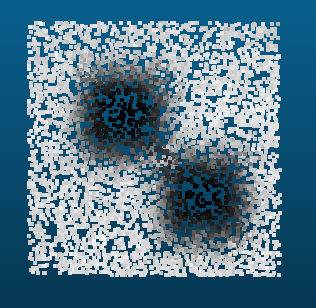}
    }
    \hfill
    \subfigure[After binarisation]{
        \includegraphics[width=0.45\linewidth]{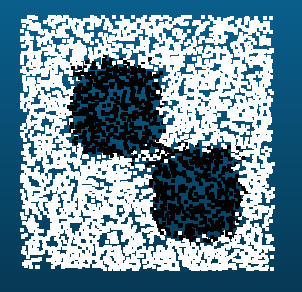}
    }
    \caption{Effect of binarisation on point cloud processing using Otsu's method}
    \label{fig:binarisation}
\end{figure}

\begin{figure*}[h]
\minipage{0.48\textwidth}
  \includegraphics[width=\linewidth]{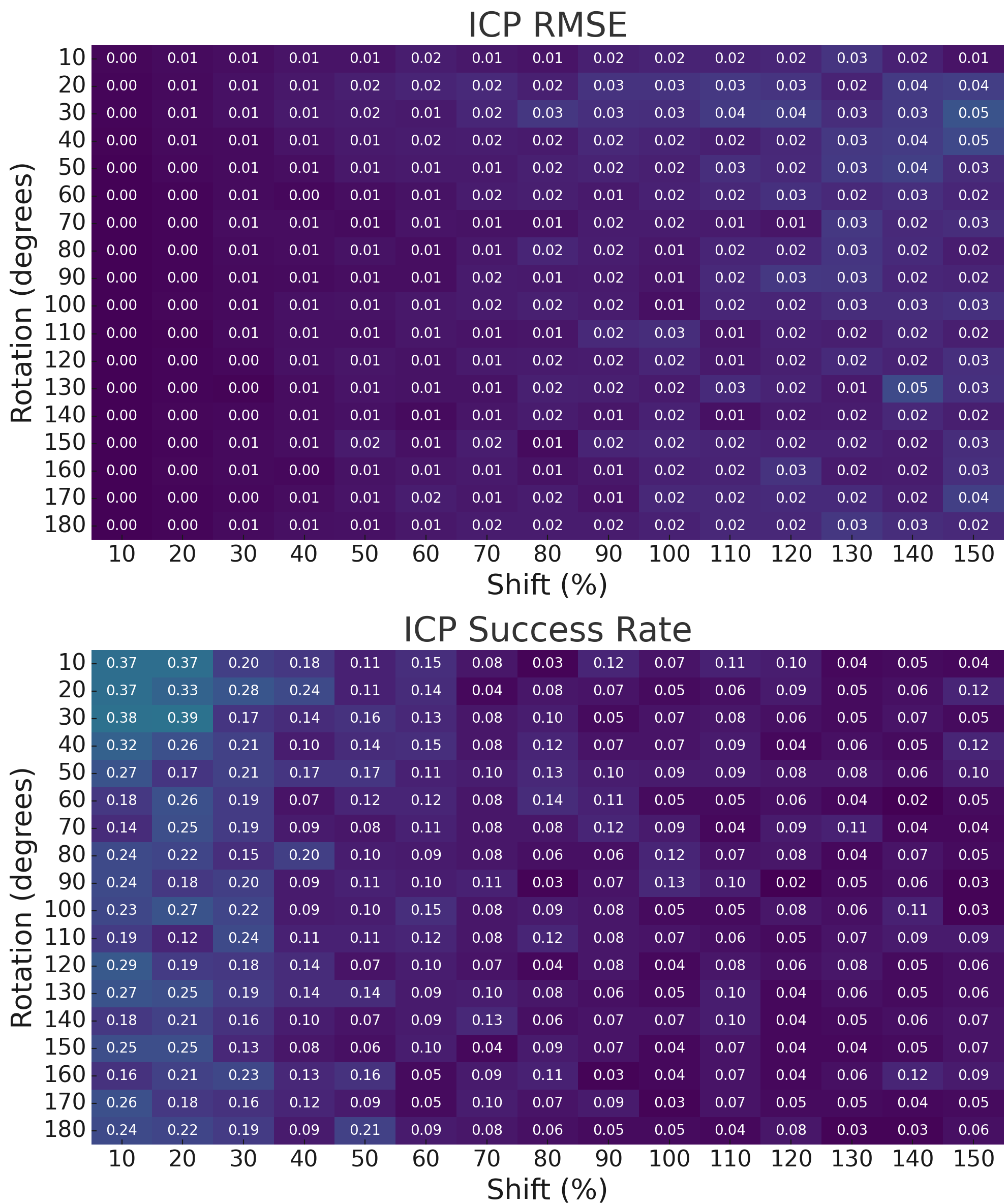}
\endminipage\hfill
\minipage{0.48\textwidth}
  \includegraphics[width=\linewidth]{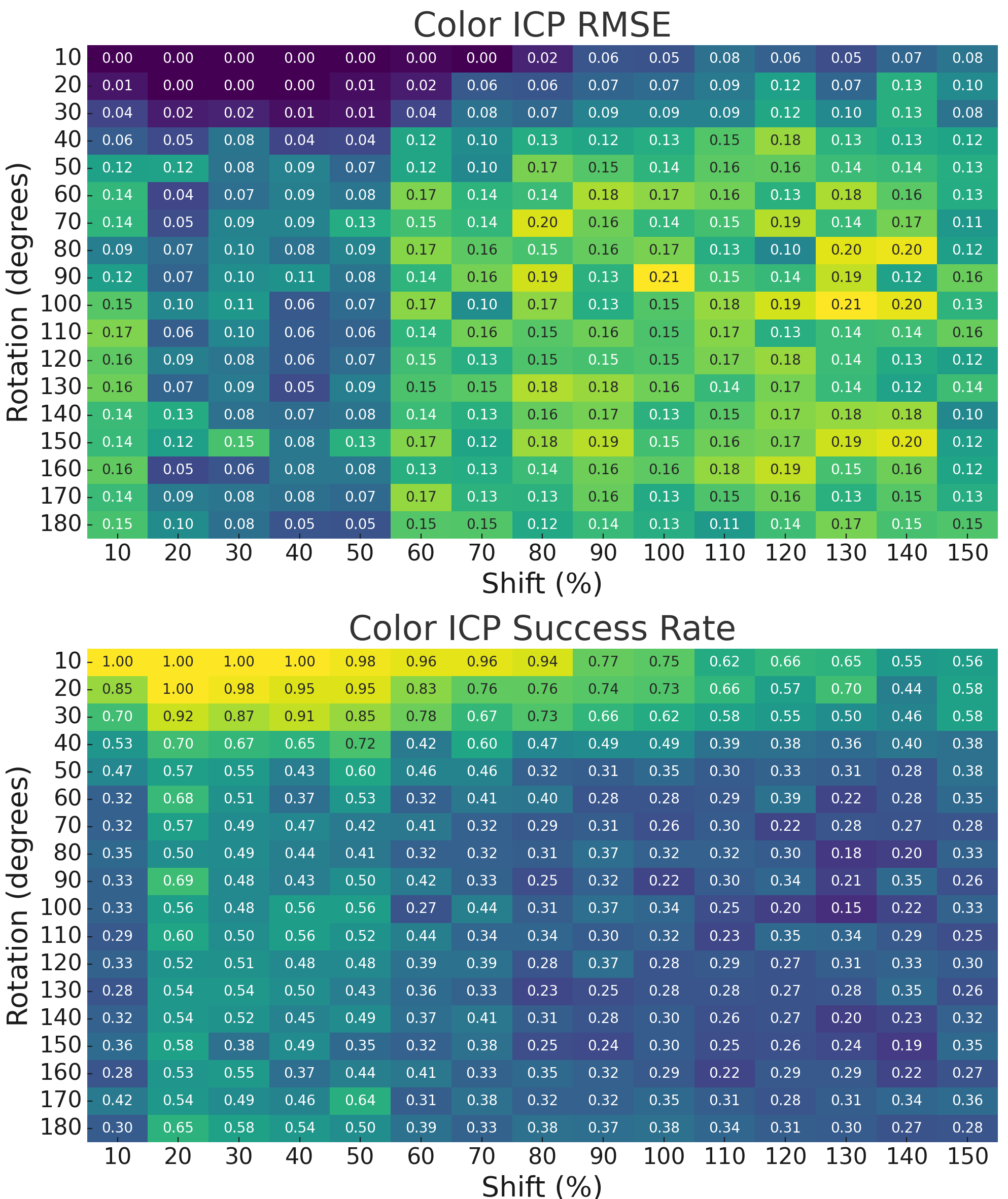}
\endminipage\hfill

\caption{Comparison of ICP and Coloured ICP results across different shifts and rotations. The results highlight the RMSE and success rate differences between the two methods under varying conditions.}
\label{fig:both_shift_rotation_results}
\end{figure*}

The experiments on simulated data will be followed up with tests on real LIDAR data captured with a Livox MID-360 sensor. Figure \ref{fig:target_visual} shows a visual representation of the problem we are tackling, with the rightmost scan being the most challenging situation. 

\textbf{Real Data} In order to evaluate the proposed checkerboard registration method in real-world scenarios, we collected indoor point cloud data using our Livox LiDAR sensors. The scanned scene includes a physical checkerboard composed of black and white squares. As we are not aware of commercial reference software that can be applied to arbitrary unordered point clouds, we do not have ground truth values for this experiment. Therefore, we conduct a second experiment on a dataset acquired of the same scene with a Leica RTC 360 survey-grade terrestrial laser scanner. We use vendor-provided software Leica Register360 to measure the centres of the targets in the scene. This provides ground truth values, albeit on an ordered and rather noise-free point cloud.

We first isolate the region of interest from the raw LiDAR data by cropping a bounding box around the checkerboard's assumed initial location. The synthetic checkerboard template is initialised near the assumed position of the real checkerboard. To match the scale and geometry of the real chequerboard, we resized the synthetic model to the a priori known physical dimensions of the realcheckerboard. In order to leverage colour-based registration, we convert intensity information of the synthetic checkerboard into corresponding RGB values. Following the extraction of the real checkerboard and the preparation of the synthetic one, we perform a first alignment step using Point-to-Plane Iterative Closest Point (ICP). 

Due to the nature of LiDAR sensing, black surfaces generally exhibit lower reflectivity compared to white surfaces, often leading to weaker return signals and higher measurement noise. In our Livox sensor data, the black squares on the checkerboard display significantly more noise and outlier points than the white squares. There are already previous reports on the issues around the intensity values and effects on ranging accuracy of this category of LiDAR sensors \citep{zhang_investigation_2023}.

To mitigate the impact of the noise on point cloud registration, we apply a RANSAC-based outlier removal step on the extracted real checkerboard. This is to eliminate outliers, predominantly over the black squares. 

It is well known from template matching in images that differences in colour or grey values between source and target need to be considered. For example, in least-squares matching this is often achieved with estimating an offset and scale factor for the intensity. However, in \citep{park_colored_2017} the photometric cost function is formulated as the squared differences of intensity values. We must therefore adapt the intensities before Coloured ICP is applied.




To address these problems, we perform a binarisation of the point cloud intensities using Otsu’s method. This binarised representation facilitated the identification of corresponding points during Coloured ICP, ensuring more accurate and robust registration by reducing ambiguity in grayscale intensity variations.

With the pre-processed real checkerboard (noise reduction and binarisation) and the synthetic checkerboard, we execute Coloured ICP to further refine the alignment. The full workflow with example data is shown in Figure \ref{fig:Registration_visualisation}.

\section{Results}\label{RESULTS}
\begin{figure}[t]
    \subfigure[RMSE of Coloured ICP ]{
        \includegraphics[width=1\columnwidth]{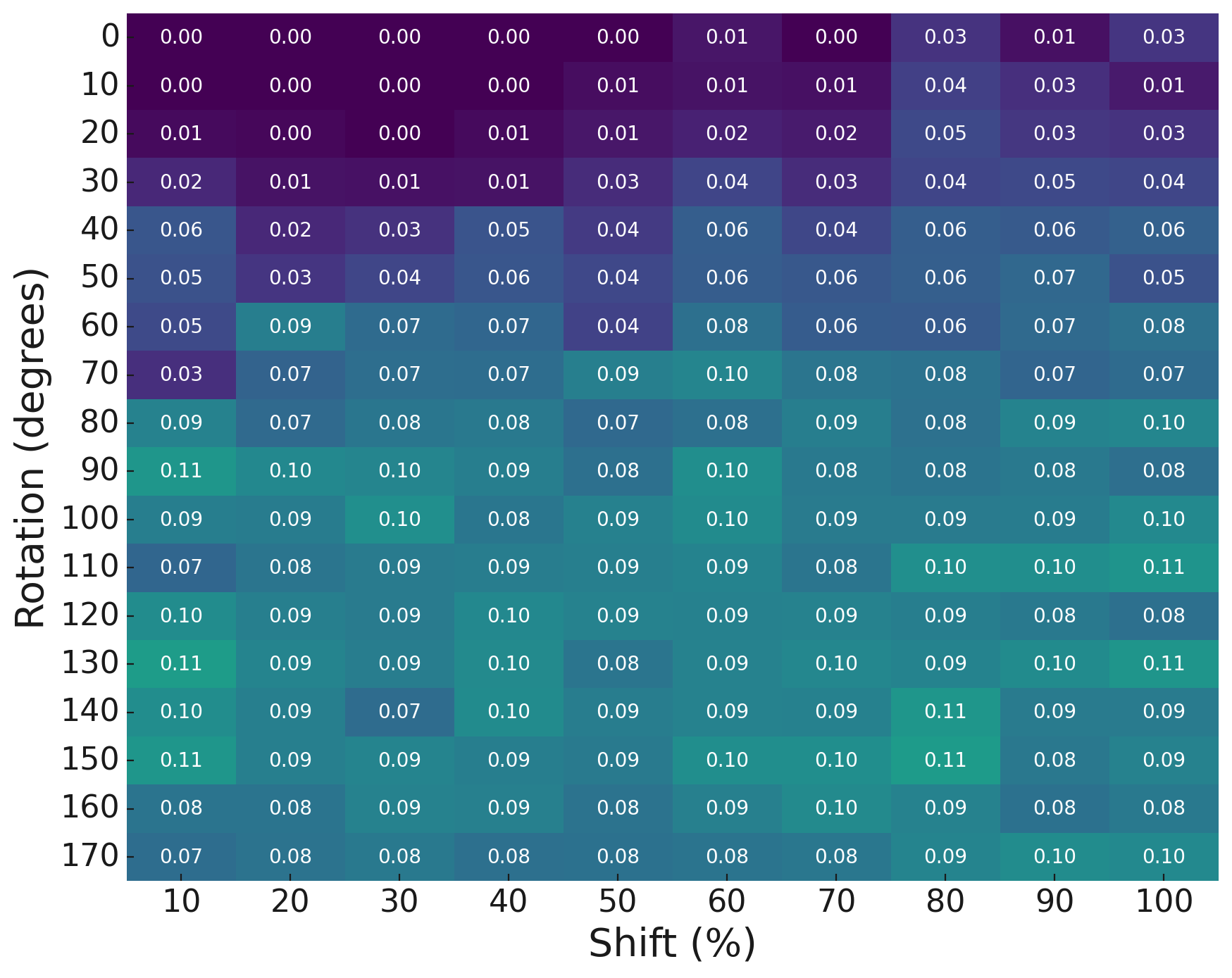}
    }
    \subfigure[Success Rate of Coloured ICP ]{
        \includegraphics[width=1\columnwidth]{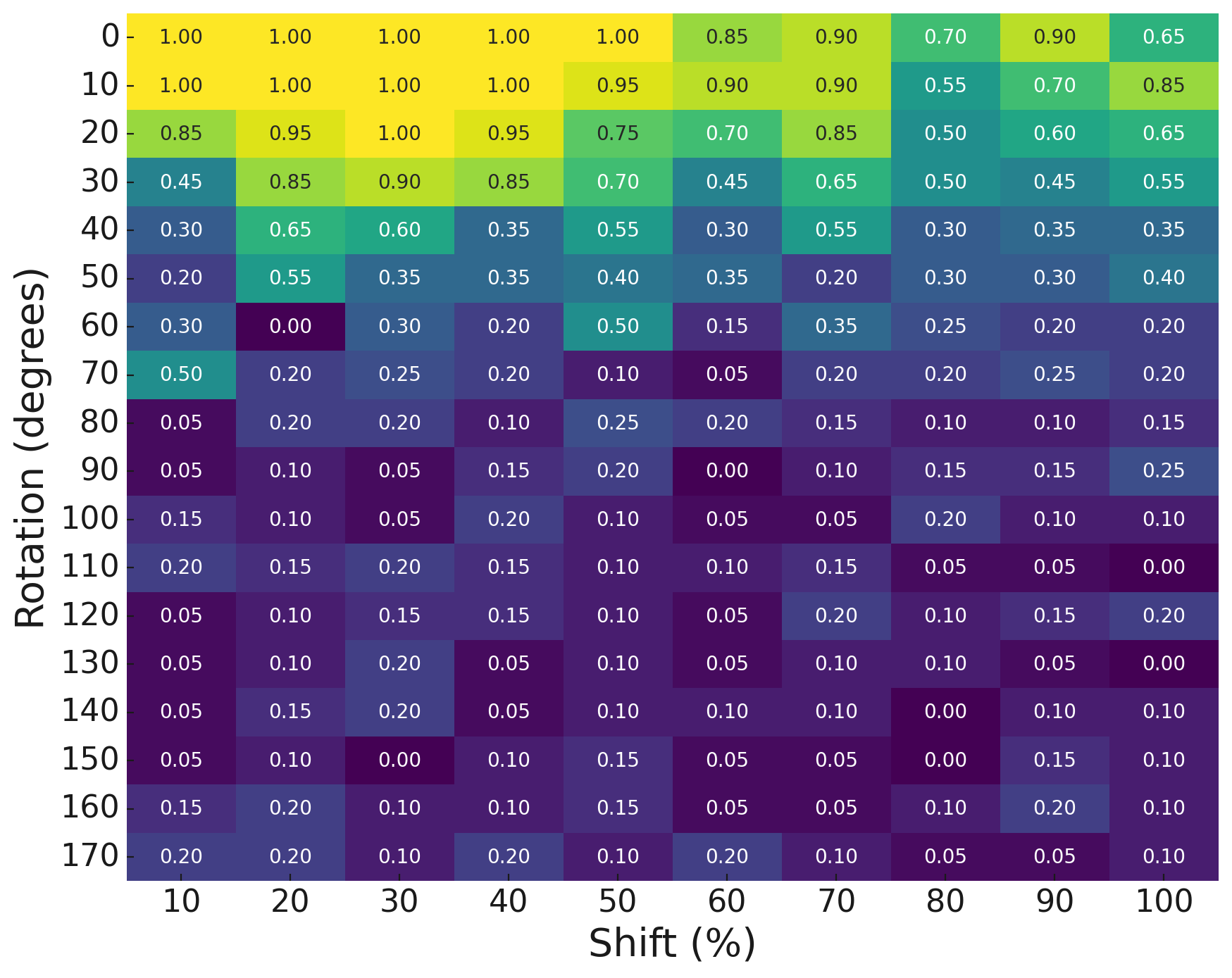}
    }

    \caption{Results of Coloured ICP with noise in the target data. (a) RMSE across different shifts and rotations.(lower values indicate better performance) (b) Success Rate across different shifts and rotations.(higher values indicate better performance)}
    \label{fig:both_shift_rotation_noise_results}
\end{figure}

\begin{figure}[!ht]
    \subfigure[RMSE of ICP and Color ICP]{
        \includegraphics[width=1\columnwidth]{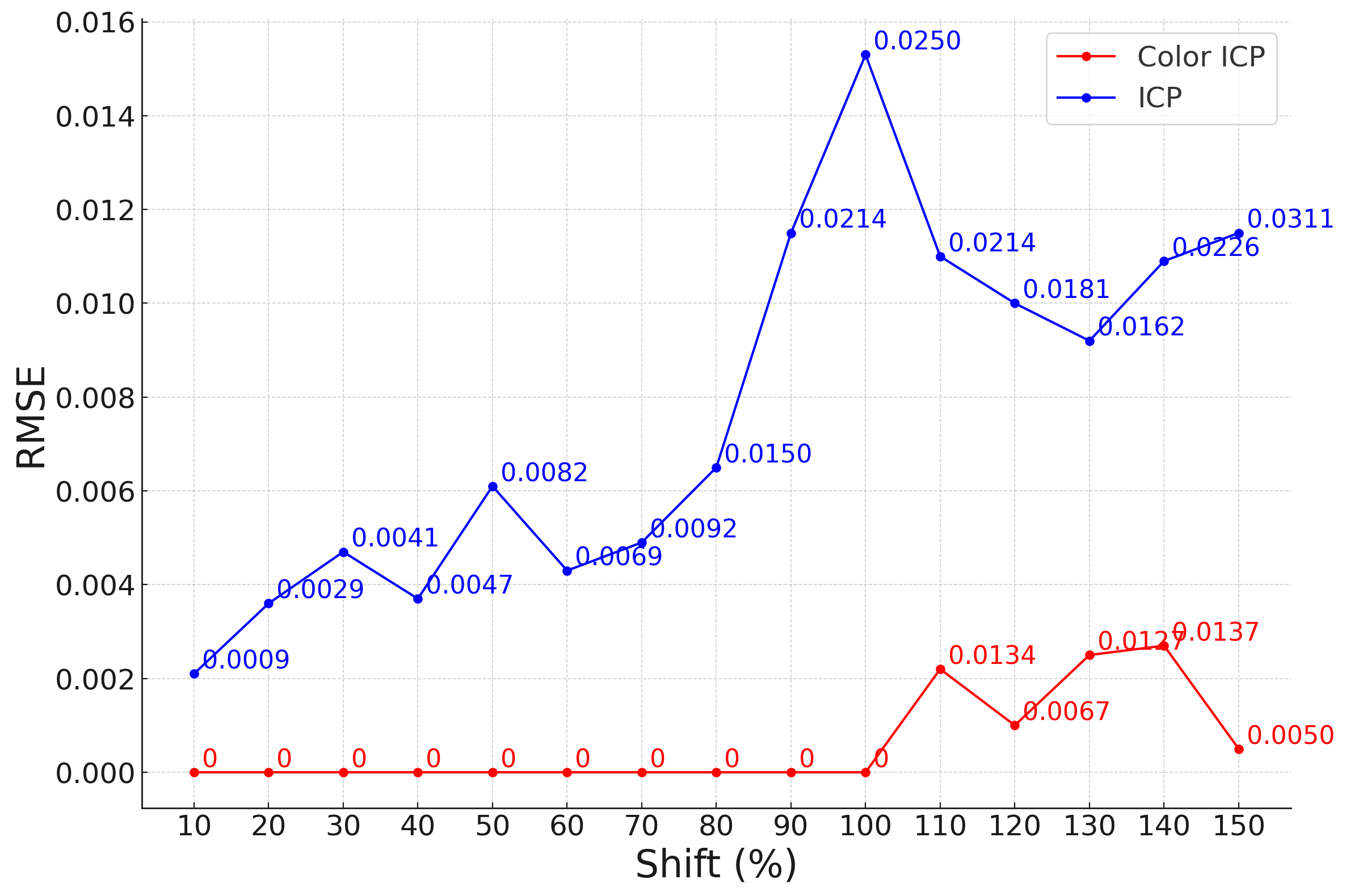}
    }
    \subfigure[Success Rate of ICP and Color ICP]{
        \includegraphics[width=1\columnwidth]{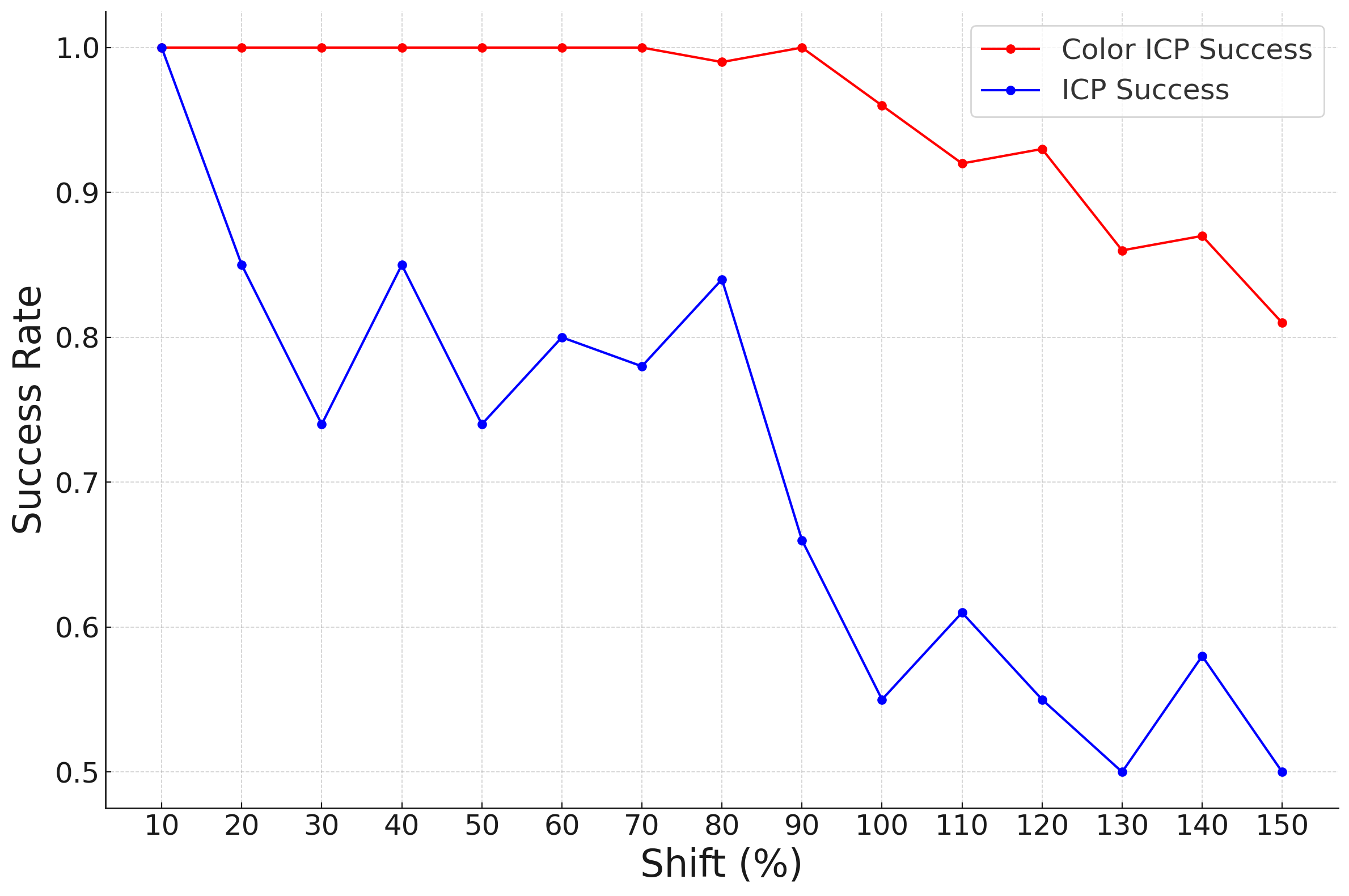}
    }

    \caption{(a) RMSE of ICP and Coloured ICP, with numbers next to each point representing the standard deviation from 100 execution counts. (b) Success Rate of ICP and Coloured ICP, comparing their performance across different shift percentages.}
    \label{fig:only_shift_results}
\end{figure}

We start by conducting experiments on synthetic data. We create the source template as a regular grid of points with the colouring of a chequerboard. In the initial experiments, the same data act as the target point cloud to be matched against. Figure \ref{fig:only_shift_results} shows that the coloured ICP is able to maintain a high success rate and a low RMSE for shifts up to 90\% when only shifts between the source point cloud and target point cloud are considered. We show pure ICP as a baseline here to reiterate that pure ICP cannot reliably find a match for planar targets.

In a second round of experiments for synthetic data, we combine shift and rotation. Figure \ref{fig:both_shift_rotation_results} shows that adding rotation reduces the success rate. For rotations up to 30 degrees, shifts up to 60\% still provide a better success rate than 70\%. Figure \ref{fig:both_shift_rotation_noise_results} shows the performance when noise is added to the target point cloud. The range for successful convergence is slightly reduced.

%

\begin{figure}[h]
    \centering
    \subfigure[Without pre-processing]{
        \includegraphics[width=0.45\linewidth]{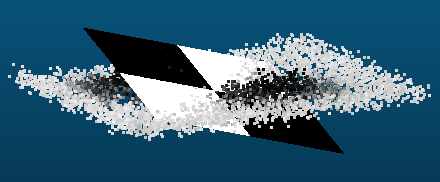}
    }
    \hfill
    \subfigure[With pre-processing]{
        \includegraphics[width=0.45\linewidth]{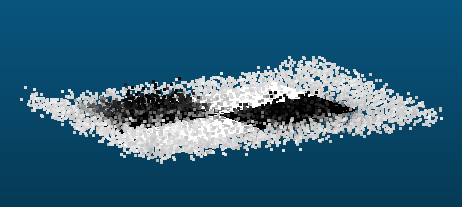}
    }
    \caption{Effect of pre-processing on Coloured ICP registration.}
    \label{fig:pre_processing_effect}
\end{figure}

\begin{figure}[h]
    \centering
    \subfigure[front]{
        \includegraphics[width=0.45\linewidth]{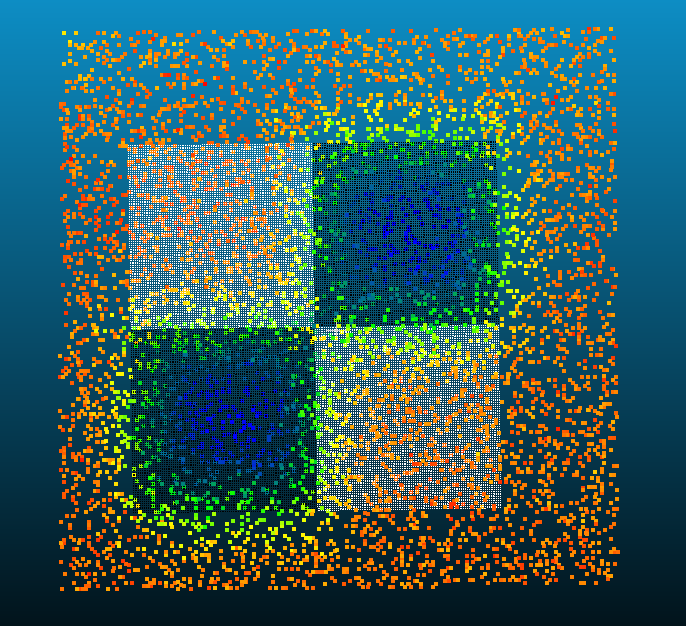}
    }
    \hfill
    \subfigure[side]{
        \includegraphics[width=0.45\linewidth]{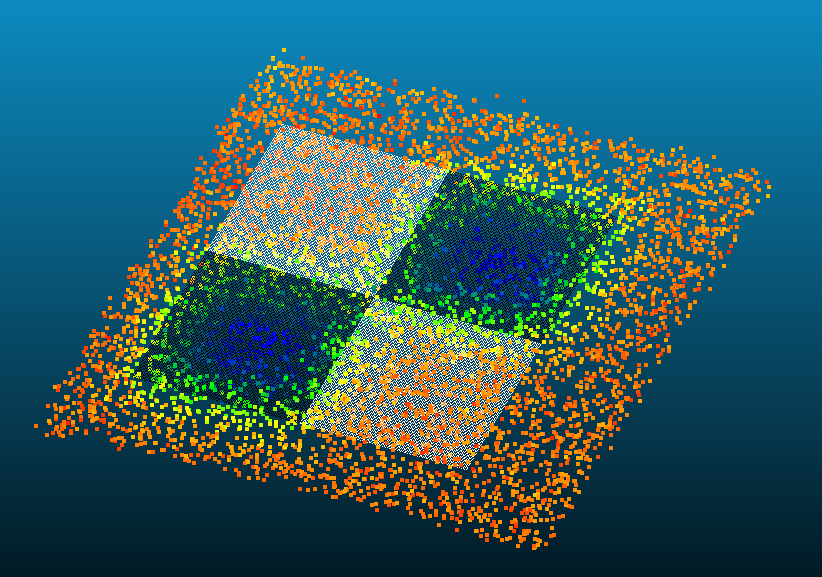}
    }
    \caption{Final alignment between template and scanned target from a low-cost sensor. We use pseudo colour for visualising the sensor data to better see the intensity alignment. }
    \label{fig:final_alignment_psudo_colour}
\end{figure}

The next set of experiments are conducted on real point clouds that were acquired over a test field with different checkerboard targets using both a survey-grade terrestrial laser scanner and a low-cost automotive scanner. For the low-cost system, we have no ground truth data, so we will provide qualitative results. Figure \ref{fig:ransac_before_after} shows that the RANSAC filter has effectively removed the outliers while preserving most of the checkerboard points. Figure \ref{fig:binarisation} (a) shows the 'bleeding' across the black and white edges from the low-cost system. Binarisation shown in Figure \ref{fig:binarisation} (b) generates sharper edges which has improved results in our experiments. Figure \ref{fig:pre_processing_effect} illustrates how pre-processing impacts the outcome. Without preprocessing, the results are noticeably more susceptible to noise and exhibit a tilt. In  Figure \ref{fig:final_alignment_psudo_colour}, we show the final alignment that is achieved which gives us the centre of the target via the transformation parameters obtained with the Coloured ICP.

The experiments on the survey-grade data are performed to validate the approach and allow us to compare the results to vendor-calculated reference values which we consider as ground truth values. We use a Leica RTC 360 laser scanner. This scanner is specified by the vendor with a 3D accuracy of 1.9 millimetre at 10 metre range. We used the vendor provided software Leica Register 360 to calculate reference values for the target centres. We achieved an average accuracy on the target centres of 1.1 millimetre with our approach.

\section{Conclusion}\label{CONCLUSION}
The primary contribution of this research is the systematic evaluation of key parameters that affect the accuracy of the Coloured ICP algorithm to measure checkerboard centres in unordered point clouds. We explored parameters such as Gaussian noise, translation, and rotation in a controlled setup using synthetic data. The finalised settings are validated using real scanned datasets, demonstrating that the approach effectively measures target centres. When compared to reference methods the approach delivers the expected accuracies. The approach is capable of handling unordered point clouds in the presence of significant noise both in range and intensity.

The experiments have also raised some new questions around the quality of the low-cost scanner data. At the moment, we do not know enough about the physical or optical properties of the sensor. For example, the 'bleeding' that is visible in Figure \ref{fig:binarisation} (a) could be caused by the spot size of the scanner or a cross-talk on the detector between subsequent point acquisitions. Likewise, the interaction of reflectance and range measurement is a continuing issue. Further investigations would be beneficial to create a better model of the sensor's characteristics. This would also aid the simulation of the sensor and could provide more realistic synthetic data.

\section{Acknowledgment}
June Moh Goo is supported by the Engineering and Physical Sciences Research Council through an industrial CASE studentship with Ordnance Survey\\ (Grant number EP/X524840/1).

{
	\begin{spacing}{1.17}
		\normalsize
		\bibliography{ISPRSguidelines_authors} 
	\end{spacing}
}

\end{document}